\documentstyle[editedvolume,psfig]{crckapb}

\begin{opening}
\title{CHEMICAL EVOLUTION MODELS \protect\\
       ALONG THE HUBBLE SEQUENCE}

\author{M. MOLLA}
\author{A.I. DIAZ}
\institute{Universidad Aut\'{o}noma de Madrid,
       28049 Cantoblanco, Spain}
\author{F. FERRINI}
\institute{Universit\`{a} di Pisa, Piazza Torricelli 2, 56100 Pisa, Italy}

\end{opening}

\runningtitle{CHEMICAL EVOLUTION MODELS ALONG THE HUBBLE SEQUENCE}

\begin{document}
\begin{abstract}

We show a grid of multiphase models whose results are valid for any
 spiral galaxy, using as input the luminosity or the rotation velocity
 and the morphological type, measured by the classical index T from 1 to 10.

\end{abstract}

\section{Introduction}

A large number of numerical chemical evolution models has been
developed, in order to relax the I.R.A. (Instantaneous Recycling
Approximation) used in the first analytical models (Tinsley, 1980;
Clayton, 1984). Different hypotheses about initial conditions, and
Star Formation Rate (SFR) and Initial Mass Function (IMF) laws have
been assumed. Most of them include a gas infall onto the disk, since
the observed radial gradient of abundances in spiral disks (Milky Way
Galaxy --MWG-- among them), can not be explained by a closed box
model. Most of works have only been applied to MWG.  Exception
to this is our chemical multiphase model, which we also used for a
sample of external spirals (Moll\'{a} et al. 1996; 1999). On the other
hand, most of models being numerical, results are shown in a graphical
way, making difficult their possible use for the scientific community
not directly involved in the chemical evolution field.

We have computed a grid of chemical multiphase evolution models, which
allow to compare galaxy observations with theoretical models for a
large range of possibilities in luminosities and morphological types
and whose results will be available for the astronomical community.

\section{The Multiphase Model Description and Previous Results}

We assume a spherical protogalaxy with a gas mass which
collapses to fall in the equatorial plane by forming the disk as a
secondary structure. The infall rate of gas from a halo region to the
disk is proportional to a parameter $f$.  The sphere is divided into
concentric cylindrical regions 1 kpc wide, with a {\sl halo} and a
{\sl disk} region.  Stars form out in the halo, by a Schmidt law with
n$=1.5$ and a proportionality factor $K$.  In the disk stars form in
two steps: molecular clouds form from the diffuse gas by a Schmidt
law, n$=1.5$, with a proportionality factor $\mu$; then cloud-cloud
collisions produce stars by a {\sl spontaneous} process, at a rate
proportional to a parameter $h$.  Moreover a {\sl stimulated} star
formation process, proportional to a parameter $a$, is assumed, by the
interaction of massive stars with surrounding molecular clouds.

The model for the Solar Neighborhood --SN-- (Ferrini et
al. 1992) allowed to estimate the values of parameters $f$, $K$, $\mu$,
$H$ and $a$ in this region, in reproducing a large number of
observational constraints. Then the model was applied to the whole
galactic disk in Ferrini et al. (1994), where we calculated the radial
variation of parameters through the use of process efficiencies
($\epsilon$'s). These $\epsilon$'s were calibrated from the SN model
and assumed to remain constant for the whole MWG.  Resulting radial
distributions are in agreement with data, in particular for diffuse
and molecular gas densities.

The infall rate $f$ for other spirals was calculated (Moll\'{a} et
al. 1996; 1999) from the collapse timescale $\tau_{0}$. By knowing
that it depends on the total mass ( Gallagher, Hunter \& Tutukov 1984)
it may be computed from its value for the MWG model, $\tau{\odot}$,
and the ratio between total masses:
$\tau_{0}=\tau_{\odot}(M_{9},gal/M_{9,MWG})^{-1/2}$.  The efficiencies
$\epsilon_{\mu}$, and $\epsilon_{h}$ changed according the Hubble
type, T. The two efficiencies $\epsilon_{K}$ and $\epsilon_{a}$ are
constant for all haloes and galaxies.  Results show good agreement
with observed radial distributions of abundances, gas and star
formation rates and with other observed correlations.

\section{The Generalization of the Model} 
\begin{figure}
\centerline{%
\psfig{file=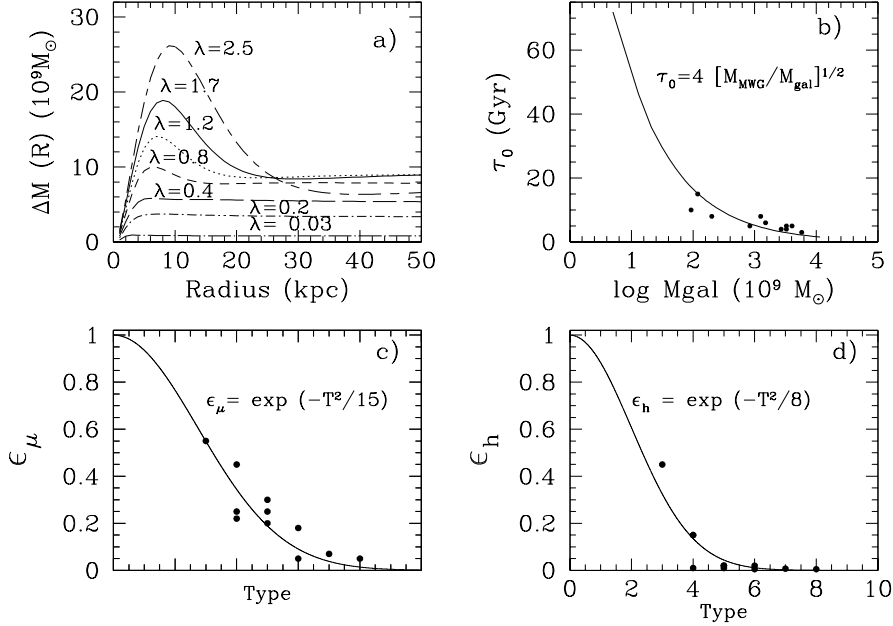,width=\textwidth,angle=-90}%
  }

\caption{Model input and parameters. a) $\Delta$ M (R) for each 
$\lambda=$L/L$_{0}$ ($L_{0}=2.5 10^{10}L_{\odot}$).
b) $\tau_{0}$ {\sl vs} the total mass of each galaxy in
logarithmic scale. 
The dot symbols are the values used in previous models.
c) Efficiencies $\epsilon_{\mu}$ {\sl vs} the morphological type T. 
d) The same for efficiencies $\epsilon_{h}$}
\label{figure1}
\end{figure}

The Universal Rotation Curve from Persic, Salucci \& Steel (1996), an
analytical expression of V(R) for any luminosity, measured by
$\lambda= L/L_{0}$, is used to calculate radial distributions of total
mass M(R), and the mass included in each one of our cylinders $\rm
\Delta M$ (R). The latter is shown in Fig.1a).  The infall rate
parameter $f$ is inversely proportional to the collapse time scale,
$\tau_{0}$, calculated with the same expression of the above section.
Values are shown in Fig.1b) {\sl vs} the total mass of each galaxy.
Points are the values used in our previous models of spiral galaxies.

Efficiencies $\epsilon_\mu$ and $\epsilon_h$ to form molecular clouds
and stars, respectively, are assumed to be dependent on the morphological
type T, as we show in Fig1.c) and 1d). A fit to the values
used in previous works has been done and included in the model, 
allowing to obtain 10 different types of galaxies for each radial
distribution of mass $\rm \Delta M(R)$.

\section{Preliminary Results and Conclusions}

\begin{figure}
\centerline{%
\psfig{file=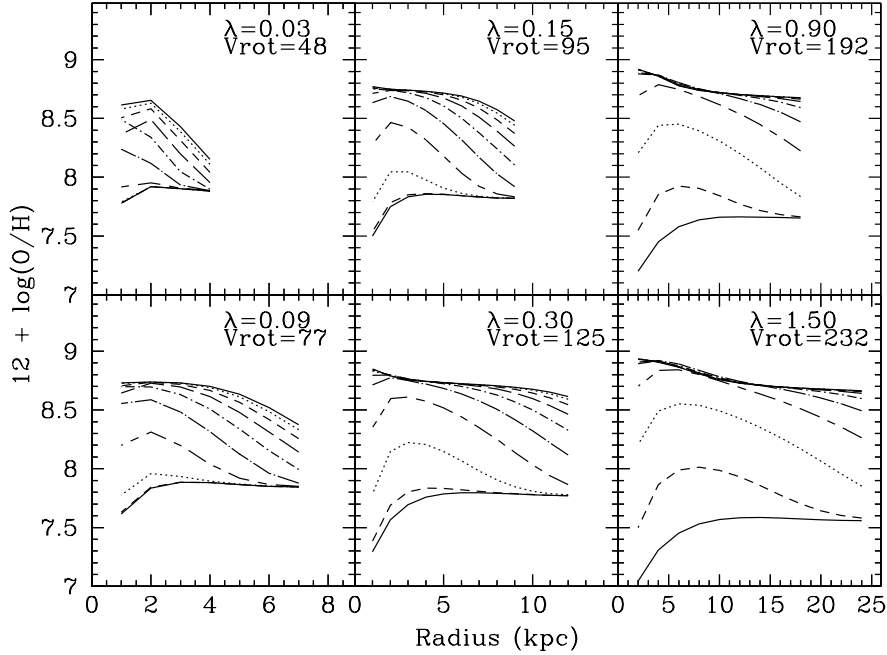,width=\textwidth,angle=-90}%
  }

\caption{Radial distribution of oxygen abundances $\rm 12 + \log{(O/H)}$
for different $\lambda$'s values. In each panel 10 different 
types T are represented, the later in the bottom and the earlier in the top of
each graph} 
\label{figure2}
\end{figure}

A biparametric grid of models is obtained for types 1 to 10 and different
luminosities. Among the radial distributions predicted for these
theoretical galaxies, we show the ones corresponding to the
 oxygen abundances in Fig.2. 
 A radial gradient for the intermediate types ($7 < \rm{T} < 4$) appears
for all luminosities, larger  for the less massive
galaxies.  The earlier types T$ \leq 5$ reach soon a saturation
level, flattening the gradient, while the latest ones (T $ \ge 8$)
have not developed a gradient in a Hubble time, although abundances
are $12 + \log{(O/H)} \sim$ 7.5-8.

This important result reproduces the observations and it solves the
apparent inconsistency of the largest gradients appearing in late type
galaxies while some irregulars shows no gradient at all, with very
uniform abundances (see Moll\'{a} \& Roy, 1999 and references
therein). The explanation resides in the stimulated star formation
which does not depend either on the Hubble type or on the
galactocentric radius, maintaining a minimum level of star formation
constant for all regions.  Possible correlations must be still
analyze, and will be prsented in the future.

\end{document}